\documentclass[aps,prb,twocolumn, superscriptaddress,amsmath]{revtex4}

\usepackage[latin1]{inputenc}
\usepackage{graphicx}
\usepackage{bm}
\usepackage{amssymb}
\usepackage{longtable}
\usepackage{dcolumn}

\input cyracc.def
\DeclareFontFamily{U}{russian}{}
\DeclareFontShape{U}{russian}{m}{n}
        { <5><6> wncyr5
        <7><8><9> wncyr7
        <10><10.95><12><14.4><17.28><20.74><24.88> wncyr10 }{}
\newcommand{\textcyr}[1]{%
  {\fontencoding{U}\fontfamily{russian}\selectfont#1}}

\newcommand {\etal} {\textit{et al.}}
\newcommand {\Sm}[1] {Sm$^{#1+}$}
\newcommand {\SmYS}[2] {Sm$_{#1}$Y$_{#2}$S}
\newcommand {\FF} {${^7F_0} \rightarrow {^7F_1}$}
\newcommand {\HH} {${^6H_{5/2}} \rightarrow {^{6}H_{7/2}}$}
\newcommand {\BG} {\hbox{$B$-$G$}}

\begin{document}

\title{Magnetic spectral response and lattice properties in mixed-valence \SmYS{1-x}{x} solid solutions studied with x-ray diffraction, x-ray absorption spectroscopy, and inelastic neutron scattering}




\author{P. A. Alekseev}
\affiliation{LNSR, ISSSP, Russian Research Centre ``Kurchatov Institute'', 123182 Moscow (Russia)}

\author{J.-M. Mignot}
\affiliation{Laboratoire L\'eon Brillouin, CEA-CNRS, CEA/Saclay, 91191 Gif sur Yvette (France)}

\author{E. V. Nefeodova}
\author{K. S. Nemkovski}
\author{V. N. Lazukov}
\author{N. N. Tiden}
\affiliation{LNSR, ISSSP, Russian Research Centre ``Kurchatov Institute'', 123182 Moscow (Russia)}

\author{A. P. Menushenkov}
\author{R. V. Chernikov}
\affiliation{Moscow Engineering Physics Institute (State University), Kashirskoe shosse, 31, 115409 Moscow (Russia)}

\author{K. V. Klementiev}
\affiliation{HASYLAB at DESY, Notkestr.\ 85, D-22603 Hamburg (Germany)}

\author{A. Ochiai}
\affiliation{Center for Low Temperature Science, Tohoku University, Sendai 980 (Japan)}

\author{A. V. Golubkov}
\affiliation{A.\ F.\ Ioffe PTI, 194021 St.\ Peterburg (Russia)}

\author{R. I. Bewley}
\affiliation{ISIS, RAL, Didcot, Oxon, OX110QX (UK)}

\author{A. V. Rybina}
\author{I. P. Sadikov}
\affiliation{LNSR, ISSSP, Russian Research Centre ``Kurchatov Institute'', 123182 Moscow (Russia)}

\date{\today}


\begin{abstract}

Mixed-valence phenomena occurring in the ``black'' ($B$) and ``gold'' ($G$) phases of \SmYS{1-x}{x} have been studied by x-ray diffraction, x-ray absorption spectroscopy, and inelastic neutron scattering. Lattice-constant and phonon-dispersion results confirm that the valence instability occurs already inside the $B$ phase. On the other hand, pronounced temperature anomalies in the thermal expansion $\alpha(T)$, as well as in the Sm mean-square displacements denote the onset of the \BG\ transition for the compositions $x = 0.33$ and 0.45. It is argued that these anomalies primarily denote an effect of electron-phonon coupling. The magnetic spectral response, measured on both powder and single crystals, is dominated by the \Sm{2} spin-orbit component close to 36 meV. A strongly overdamped \Sm{3} contribution appears only for $x \ge 0.33$ near room-temperature. The quasielastic signal is strongly suppressed below 70 K, reflecting the formation of the singlet mixed-valence ground state. Quite remarkably, the signal around 36 meV is found, from the single-crystal spectra, to arise from \textit{two} distinct, dispersive, interacting branches. The lower peak, confirmed to exist from $x = 0.17$ to $x = 0.33$ at least, is tentatively ascribed to an excitation specific to the mixed-valence regime, reminiscent of the ``exciton'' peak reported previously for SmB$_6$.

\end{abstract}

\pacs{}



\maketitle


\section{Introduction}

Kondo insulators (KIs) form a class of $f$- (or, for a few of them, $d$-) electron systems which, near room temperature (RT), typically behave as poor metals, with conduction electrons scattering incoherently off the magnetic centers, very much in the same way as in metallic heavy-fermions. But contrary to what happens in the latter materials, this regime of incoherent Kondo-like scattering evolves gradually upon cooling to a semiconducting state, due to the opening of a narrow energy gap in the electronic density of states below a characteristic temperature of the order of 50--100 K. This peculiar behavior was taken to indicate a strong renormalization of quasiparticle states due to electron-electron correlations. For this reason, there has been considerable debate \cite{Aeppli92, Aeppli93} as to whether KIs can be classified as a band semiconductors or as Mott insulators, and to what extent their properties can be encompassed within the scope of existing theories, in particular the periodic Kondo or Anderson models. The spin dynamics in these systems is of primary interest because the latter models imply a direct interplay between charge and magnetic degrees of freedom and, in some cases, can make quantitative predictions as to, for instance, the ratio between a ``spin gap'' observed in the magnetic spectral response and the ``charge'' gap in the electronic band structure. Inelastic neutron scattering (INS) studies have now been reported for a number of KI materials. The results suggest that one should distinguish between systems based on magnetic ions with only one unpaired electron (Ce) or hole (Yb) on the 4$f$ shell, and those with a more complex configuration such as Sm or Eu. In the following, we will focus on case of samarium, for which detailed experiments performed on low-absorption isotopically enriched single crystals of mixed-valence (MV) SmB$_6$ have revealed a number of unexpected features.\cite{Alekseev95_JETP, Alekseev95_JPCM} In particular, the magnetic response measured at $T = 2$ K shows, in addition to two broad peaks reminiscent of the lower single-ion spin-orbit transitions in the \Sm{2} (4$f^6$) and \Sm{3} (4$f^5$) electron configurations, respectively, a very narrow excitation at low energy ($\hbar\omega \approx 14$ meV) with unusually steep $Q$ dependence. The latter peak is a genuine feature of the KI state, and is strongly suppressed upon heating to only 50 K. These properties have been explained\cite{Kikoin90, Mishch91, Kikoin95} by the formation of a quantum-mechanically mixed local ground state at each Sm site, due to the hybridization of the 4$f$ electrons with the $p$ orbitals of the nearest boron atoms. In such an approach, the MV ground state has an essentially local character and can be regarded as a sort of ``magnetic exciton''.\footnote{The same assumption is also at the basis of the mechanism proposed by Kasuya [T. Kasuya, J. Phys. Soc. Jpn. 65, 2548 (1996)], in which the \hbox{$d$-electron} state involved in the valence fluctuation is localized around each Sm$^{3+}$ site, and the lowest excited state is an \hbox{$s$-wave} exciton.} It can be contrasted with the itinerant approach underlying the well-known ``coherent hybridization gap'' picture.\cite{Martin79, Risebg92}

Considering that, in Sm-based materials, both fluctuating configurations \Sm{2} and \Sm{3} have a nonzero number of electrons (holes) in the $f$ shell, one may expect exchange interactions between rare-earth (RE) ions to play a more important role in the formation of the MV ground state than in Ce or Yb. Such effects, however, have not been found in SmB$_6$, possibly because of the marked local character of the MV state in this system. The question of the influence of intersite exchange on the ground state of MV semiconductors in general has remained poorly documented from the experimental viewpoint, with a few exceptions, such as TmSe. \cite{Shapiro82, Holl-Mor83} Among the Sm compounds, one promising candidate for studying this problem is SmS. SmS is an archetype unstable-valence compound which undergoes a first-order transition at room temperature from a divalent semiconducting state at $P = 0$ to a homogeneous MV state above a critical pressure of $\approx 0.65$ GPa. This transition is accompanied by a volume collapes, and by a change in the sample color from black ($B$) to gold ($G$) reflecting the metallic conductivity in the MV phase. Penney and Holtzberg\cite{Penney75} have shown that the $G$ phase can also be obtained at ambient pressure and RT by substituting of Y on the Sm site. In the \SmYS{1-x}{x} crystals they measured, the transition was found to take place when the yttrium concentration exceeds about 15\%, and the lattice constant falls below $\simeq 5.70$ \AA. This critical value, $a_c$, slightly decreases at low temperature ($\approx 5.67$ \AA\ at $T = 80$ K). Interestingly, the concentration dependence of the lattice constant indicates that, upon substitution of trivalent Y, the Sm valence starts to deviate from 2+ \textit{before} the critical concentration, $x_c$, is reached. This is in strong contrast with pure SmS in which Sm remains divalent all the way up to the critical pressure.

In $B$-phase SmS at ambient pressure, the magnetic excitation spectrum at low temperature is associated with the spin-orbit (SO) transition from the $^7F_0$ singlet to the $^7F_1$ triplet, whose free-ion energy is 36 meV. The degeneracy of the triplet is not lifted here because of the high, $O_h$, symmetry of the crystal field at the cubic RE site. It was first reported by Shapiro \etal\cite{Shapiro75} thirty years ago that the energy of this transition is \hbox{$q$-dependent} as a result of exchange interactions between \Sm{2} magnetic moments, mediated by the \hbox{$d$-band} states. This type of magnetic excitations in a system of induced magnetic moments is known for RE compounds with a singlet crystal-field (CF) ground state, where it is termed ``paramagnetic (CF) exciton''.\cite{Cooper72} The dispersive branch observed in Ref.\ \onlinecite{Shapiro75} can thus be characterized as a ``paramagnetic SO exciton''. It was shown to correspond almost exactly to the idealized singlet-triplet model. In Ref.\ \onlinecite{Shapiro75}, the exchange parameters for the first three coordination spheres of Sm were obtained from the energy dispersion using a Heisenberg model in the mean-field-random phase approximation (MF-RPA).

Several inelastic neutron scattering (INS) studies have been devoted to the phonon spectra of \SmYS{1-x}{x}, revealing a remarkable anomaly in the dispersion of longitudinal acoustic (LA) phonons in the MV $G$ phases  of both \SmYS{0.75}{0.25} under normal conditions\cite{Mook79} and SmS under pressure.\cite{Mook82} This anomaly was traced back to an electron-phonon interaction arising from a low-frequency charge-fluctuation mode specific for the MV state in samarium. The latter interpretation, based on an excitonic description of the charge-fluctuation state first suggested by Stevens,\cite{Stevens76} was elaborated in Refs.\ \onlinecite {Kikoin90} and \onlinecite{Mishch91} and successfully applied to a quantitative treatment of phonon anomalies in \SmYS{1-x}{x}, \hbox{$G$-phase} SmS, and SmB$_6$.\cite{Alekseev89} 
It then served as the starting point for a local description of the MV state in Sm,\cite{Kikoin95} as discussed at the beginning of this Section.

There has been, to date, relatively little experimental work on magnetic excitation spectra in the  \SmYS{1-x}{x} series, and a number of inconsistencies between the results have not yet been resolved. In the high-pressure experiments of McWhan \etal,\cite{McWhan78}, polycrystalline SmS was studied on both a triple-axis and a time-of-flight spectrometer. The authors report the disappearance of the magnetic signal from \Sm{2} in the MV $G$ phase. However, one can note that the measuring conditions were not optimal for the observing an excitonlike (in the sense of Ref.\ \onlinecite{Kikoin95}) inelastic peak having a steep $Q$ dependence of its intensity as found in SmB$_6$. For the solid solution \SmYS{0.75}{0.25}, powder experiments at ambient pressure\cite{Mook78, Holl-Mor88} have revealed a broad, \hbox{$Q$-dependent}, signal close to the position of the SO transition in both the $B$ and the $G$ phases. In Ref.\ \onlinecite{Holl-Mor88}, this broadening was interpreted as due to a substructure produced by the hybridization of magnetic (SO) modes with optic phonon modes. Also worth mentioning is the attempt made by Weber \etal\cite{Weber89} to study the temperature evolution of quasielastic magnetic scattering in \SmYS{0.75}{0.25}. However, their polycrystal, time-of-flight, experiment suffered from considerable phonon background, and the measurements at low temperature were hampered by a low incoming neutron energy, which precluded the analysis of the energy-loss side of the spectra. This point will be further discussed in section \ref{sec:discussion}.

In a previous paper,\cite{Alekseev02_PRB} we have reported single-crystal INS measurements on the $B$-phase alloy \SmYS{0.83}{0.17}. Two dispersive magnetic modes, interacting with each other, were observed, lending support to the idea that, contrary to the case of SmB$_6$, Sm-Sm exchange interactions play a significant role in the formation of the excitation spectra of the MV state in this system. It was argued that the two modes might correspond, respectively, to excitations from the ``parent'' \Sm{2} state and from an excitonlike state reminiscent of SmB$_6$. Further increasing the Y concentration is known to result in a transition to the $G$ phase with a higher average Sm valence. From the preceding considerations, this should lead to $i$) the suppression of the dispersion, $ii$) enhanced excitonic properties of the ground state, implying steeper $Q$- and $T$- dependences of the corresponding peak intensities, and $iii$) changes in the parameters of the higher energy, single-ionlike excitations associated with the two parent states \Sm{2} and \Sm{3}. An important feature of the excitonic MV ground state is the correlation found experimentally in hexaborides\cite{Alekseev97} between the energy of the new magnetic excitation and the average Sm valence. It is not obvious, however, that a similarly strong dependence of the excitation energy should occur in the sulfides, because $f$ electron states from a given Sm site are expected to hybridize with the $d$ states of neighboring Sm atoms and, therefore, delocalization may be not so strong as in SmB$_6$ where hybridization is thought to primarily involve the $p$ orbitals from boron.

In the following, we report a detailed INS study of \SmYS{1-x}{x} ($0 \leq x \leq 0.45$) performed on both powder and single-crystal samples, covering a wide temperature range, and including measurements under an applied pressure. We also present \hbox{x-ray} diffraction data on the temperature dependence of the lattice constant, the thermal expansion coefficient, and the Sm mean-square displacements, as well as the temperature evolution of the Sm valence obtained by \hbox{x-ray} absorption (XAS) spectroscopy. A comprehensive analysis of these different results allows us to make a connection between dynamical characteristics of the material and its valence state, and provides clear evidence for the major role played by exchange interactions in the properties of the ground state.


\section{Experiments}

 \begin{figure} [b]
	\includegraphics [width=0.70\columnwidth] {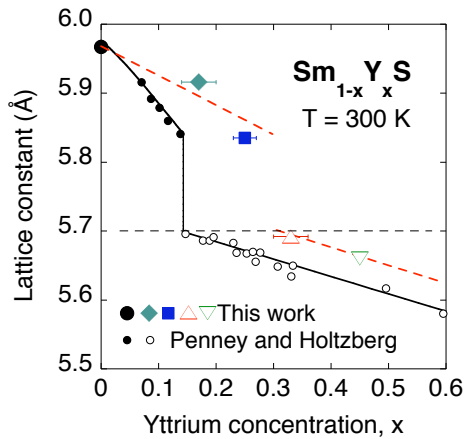}
\caption{\label{lattCnst} (Color online) Room-temperature lattice constants of the \SmYS{1-x}{x} solid solutions studied in this work (error bars for lattice constants are less than symbol size); marker types for each concentration are used consistently in the rest of the paper; closed (open) symbols denote samples which are visually in the $B$ ($G$) phase, respectively, at room temperature. Previous data from Ref.\ \onlinecite{Penney75} (small circles) are shown for comparison (see text); dashed line: experimental upper limit of the lattice constant for the existence of $G$ phase at $T = 300$ K.}
 \end{figure}

The materials used in the present experiments were prepared from $^{154}$Sm isotope with  $98.6\%$ enrichment, containing less than $0.2\%$ highly absorbing $^{149}$Sm. Samples with Y concentrations $x=0.17 \pm 0.03$ ($a$), $0.25 \pm 0.02$ ($b$), $0.33 \pm 0.03$ ($c$), and 0.45 ($d$), as well as La$_{0.75}$Y$_{0.25}$S ($e$) and pure SmS ($f$) were synthesized, by essentially identical procedures, either at Tohoku University in Sendai ($a$, $c$) or at A.~F. Ioffe PTI in St. Petersburg  ($b$, $d$, $e$, $f$). Samarium metal was first purified by the Bridgman method in a sealed tungsten crucible, then prereacted with Y, and S in a silica tube under vacuum using a conventional electric furnace. Single crystals of pure SmS and \SmYS{1-x}{x} solid solutions were grown by the Bridgman method in electron-beam-welded tungsten crucibles, then annealed for several days at $1000^\circ$C. Powder for neutron time-of-flight (TOF) measurements was obtained by crushing the smaller single-crystal pieces that were not suitable for triple-axis experiments. \SmYS{0.55}{0.45} and the reference compound La$_{0.75}$Y$_{0.25}$S were synthesized in polycrystal form because they were intended for TOF experiments only. Powder x-ray diffraction confirmed the NaCl crystal structure of all materials, as well as the absence of detectable impurity phases. The concentrations of samples $a$, $b$, and $c$ were obtained by optical spectral analysis, those given above for samples $d$ and $e$ are nominal values. Lattice parameters obtained by x-ray diffraction at room temperature are listed in Table~\ref{tabLattConst}. For the compositions $x = 0.33$ and 0.45, $a_0$ is smaller that the upper limit of $5.70$ \AA\  for the existence of the $G$ phase determined by Penney \etal\cite{Penney75} This is consistent with the gold color exhibited by crystals $c$ and $d$. On the other hand, the samples with $x = 0.17$ and 0.25 are found to be in the $B$ phase (dark-purple color), with lattice constants  in the 5.8--5.9 {\AA} range. As shown in Fig.\ \ref{lattCnst}, this implies a significant shift in the Y critical concentration in the present materials as compared to Ref.\ \onlinecite{Penney75}, which is probably due to differences in preparation procedures. In the present data, no discrepancy was found between materials from two different origins. Furthermore, the limit of $a_0 \leqslant 5.70$ {\AA} for the existence of the $G$ phase at RT is consistent with all existing results and, in the following, $a_0$ will thus be assumed to be the relevant parameter for comparing results from different studies.


\begin{table}
\caption{\label{tabLattConst}Room temperature lattice parameters $a_0$ of measured \SmYS{1-x}{x} and La$_{0.75}$Y$_{0.25}$S samples.}
\begin{ruledtabular}
\begin{tabular}{l c c c c c c}
$x$ & 0 & 0.17 & 0.25 & 0.33 & 0.45 & 0.25\footnote{La$_{1-x}$Y$_x$S}\\
$a_0$ (\AA) & 5.967(2) & 5.916(2) & 5.835(2) & 5.692(2) & 5.660(2) & 5.769(2)\\
\end{tabular}
\end{ruledtabular}
\end{table}


The temperature dependence of the lattice constants in the \SmYS{1-x}{x} solid solutions ($a$--$d$) and the nonmagnetic reference compound La$_{0.75}$Y$_{0.25}$S ($e$) have been determined by x-ray diffraction (Cu $K_{\alpha}$) in the temperature range $10 \leqslant T \leqslant 300$ K. The values were obtained by refining the positions of several high-angle reflections, and the peak intensities of the same reflections were used to derive the relative values of the mean-square displacements at the rare-earth site. 

The Sm valence for the compositions $x = 0.33$ and 0.45 was measured by $L_{\text{III}}$-edge x-ray absorption spectroscopy (XAS) on samples $c$ and $d$. The experiments were performed on the $E4$ station of the \textit{DORIS-III} storage ring at HASYLAB (DESY, Hamburg, Germany), using a Si (111) double-crystal monochromator. The energy resolution was estimated to be better than 1.5 eV at 6 keV. The measurements were performed by first cooling down the sample to 20 K, then increasing the temperature stepwise to 300 K. The method for determining the valence was described in Ref.\ \onlinecite{Tsvya02}.

Neutron powder measurements were performed on the TOF spectrometer HET (ISIS, Rutherford-Appleton Laboratory, UK). For each concentration, the total available amount of material, of about 4 g, was used to optimize counting statistics. The incoming neutron energy was $E_i = 300$ meV, making the energy range of spin-orbit excitations for both Sm$^{2+}$ and Sm$^{2+}$ accessible with sufficient resolution ($\Delta E=13$ meV at zero energy transfer, further improving in the neutron-energy-loss region).

Single-crystal inelastic neutron scattering experiments were carried out on the 2T thermal-beam triple-axis spectrometer (Laboratoire L\'eon Brillouin, France). The single crystals measured had volumes of 0.15 ($a$), 0.35 ($b$), 0.17 ($c$), and 0.42 ($f$) cubic centimeters. The crystal mosaicity, obtained from the full widths at half maximum (FWHM) of the neutron rocking curves on the 200 nuclear peak, was less than 1 degree in all cases. The samples were cooled from room temperature (RT) down to $T = 12$ K using a closed-cycle refrigerator. Spectra were recorded at a fixed final neutron energy of $E_{f} = 30.5$ meV (14.7 meV for studying the quasielastic signal), using a Cu 111, or PG 002 double-focusing monochromator, a PG 002 analyzer, and a graphite filter on the scattered beam to suppress higher orders. The resulting resolution (FWHM) at zero energy transfer was 2 meV for $E_{f} = 30.5$ meV, and 0.8 meV for $E_{f} = 14.7$ meV. The magnetic response was measured in the range of momentum transfer $1.99 \leqslant Q \leqslant 4.6$  {\AA}$^{-1}$. The temperature dependence of quasielastic magnetic scattering was studied near the Brillouin zone boundary, where this signal is well separated from acoustic phonons. A wave vector of $\mathbf{Q} = (1.45, 1.45, 1.45)$, with components slightly shifted from half-integral values, was chosen in order to avoid contamination by second order of Bragg scattering. The energy dispersion of selected phonon branches was also determined. Data for the longitudinal acoustic branch along the [111] direction are presented in Section \ref{subsec:lattprop}.

Pressure experiments up to $P = 0.48$ GPa were performed using a He-gas chamber cooled down in an ILL-type helium-flow cryostat. Measurements on a SmS single crystal ($f$) and on powdered \SmYS{0.67}{0.33} ($c$), were done on the triple-axis spectrometer 2T and the TOF spectrometer HET, respectively. In the latter case, a near doubling of the incoming neutron flux was achieved by reducing the frequency of the Fermi chopper by a factor of 2, at the cost of a loss in resolution ($\Delta E =22$ meV at zero energy transfer).

\section{Results}
\label{sec:results}


\subsection{Lattice properties}
\label{subsec:lattprop}

 \begin{figure}
	\includegraphics [width=0.80\columnwidth] {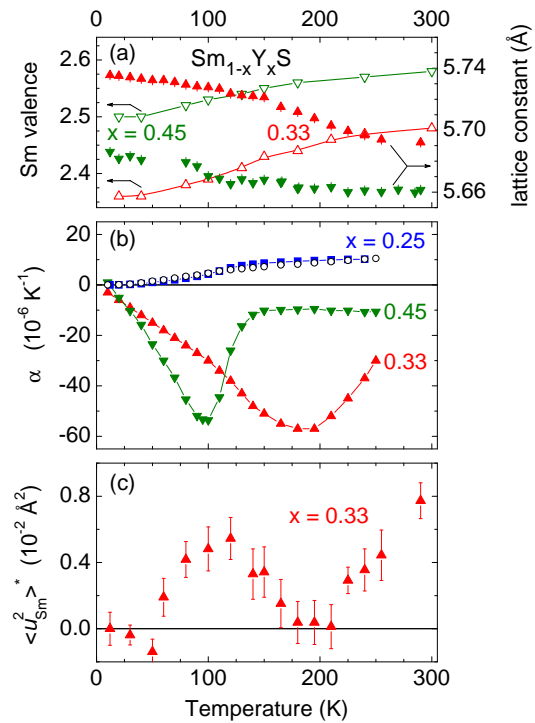}
\caption{\label{valenceThermexp} (Color online) Temperature dependences of lattice properties in \SmYS{1-x}{x} as obtained from x-ray diffraction measurements: (a) lattice constant for $x = 0.33$ and 0.45; also plotted is the valence of the same compounds obtained from XAS measurements; (b) thermal expansion coefficient for $x = 0.25$, 0.33, and 0.45; data for La$_{0.75}$Y$_{0.25}$S are plotted as open circles; (c) relative mean-square displacements of RE ions at $T = 10$ K for $x = 0.33$ (see text)}
 \end{figure}

From the temperature dependence of their lattice constants, samples $a$ and $b$ ($x = 0.17$ and 0.25) are found to be in the $B$ phase in the whole temperature range from RT to $T = 10$ K. The Sm valence in these samples\cite{Alekseev02_Pha} is almost temperature independent, $v \approx 2.19$ ($a$) and 2.22 ($b$). On the other hand samples $c$ and $d$ ($x = 0.33$ and 0.45) are in the $G$ state at RT, but undergo a phase transition at $T \approx 200$ and $\approx 100$ K, respectively, to a state with a larger lattice constant [Fig.\ \ref{valenceThermexp}(a)]. The temperatures of the \BG\ transitions quoted in the following were defined from the zero of the second derivative of $v(T)$. In Ref.\ \onlinecite{Penney75}, this effect was observed at somewhat lower Y concentrations, $0.15 \leqslant x \lesssim 0.3$ (see above), and the low-temperature ($T = 80$ K) data points were observed to lie on the extrapolation of those obtained for samples which remain black at all temperatures, suggesting that the low-temperature state is very similar, if not identical, to the $B$ phase.  Whereas the latter observation was not quantitatively confirmed in the case of our samples, one can see  in Fig.\ \ref{valenceThermexp}(a) that the electronic phase transition is indeed accompanied by a considerable decrease in the Sm valence at low temperature. Henceforth we will use the term ``$B$ phase'' indiscriminately to denote both the low-pressure state and the high-pressure, low-temperature state.\footnote{The question of whether the low-temperature, low-valence state of the collapsed samples, sometimes denoted $B'$ in the literature, should be regarded as different in nature from the original $B$ phase has been a subject of controversy. Answers may differ depending on the physical properties taken into consideration. Since there appears to be no strong case for such a distinction in the present neutron results, we chose to ignore it for simplicity.} 

The linear thermal expansion coefficients $\alpha = (1/a)(da/dT)$ of the above two alloys have been calculated from the $a(T)$ data [Fig.\ \ref{valenceThermexp}(b)]. Their temperature dependence exhibits large additional negative contributions in comparison with the nonmagnetic compound La$_{0.75}$Y$_{0.25}$S or the \hbox{$B$-phase} alloy \SmYS{0.75}{0.25}, with pronounced minima located close to the temperatures of the \hbox{$B$-$G$} phase transition. From the nuclear Bragg peak intensities, we have derived the $T$ dependence of the Sm mean square displacements, $\langle u^2_{\text{Sm}} \rangle ^{\ast}(T) = \langle u^2_{\text{Sm}} \rangle (T) - \langle u^2_{\text{Sm}} \rangle (T = \text{10}$ K) in sample $c$ ($x = 0.33$). This dependence [Fig.\ \ref{valenceThermexp}(c)] is nonmonotonic with temperature and exhibits a pronounced dip at $T \sim 200$ K, which coincides with the minimum in the thermal expansion coefficient. The connection between these two effects and the electronic phase transition is discussed in Section \ref{sec:discussion}.

 \begin{figure}
	\includegraphics [width=0.60\columnwidth] {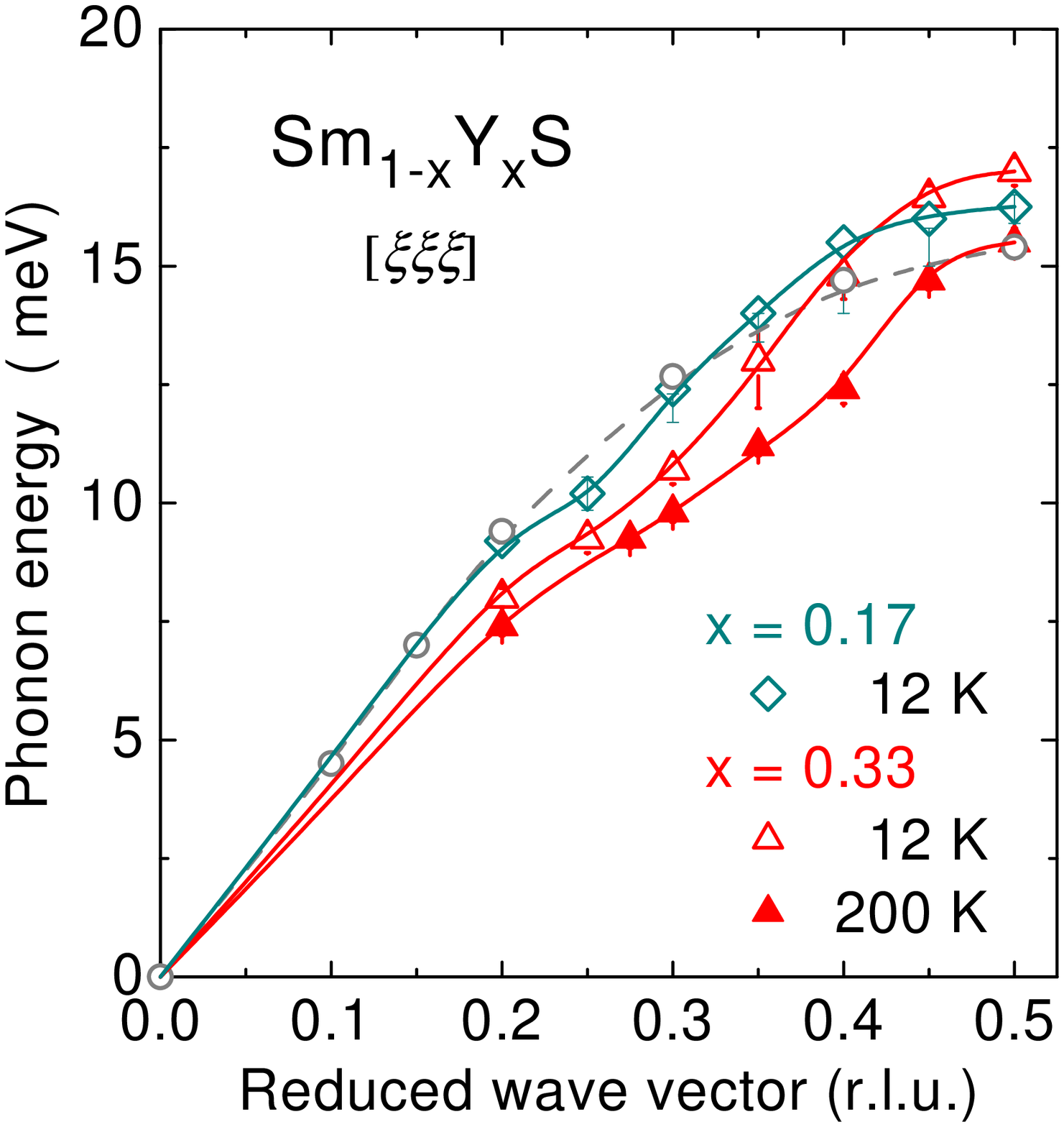}
\caption{\label{phonAnom} (Color online) LA phonon dispersion along the [111] direction for \SmYS{0.83}{0.17} and \SmYS{0.67}{0.33}; dashed line: data for SmS from Ref.\ \onlinecite{Birgeneau77}}
 \end{figure}

The dispersion curves of  LA phonons for \SmYS{0.83}{0.17} ($T = 12$ K) and \SmYS{0.67}{0.33} ($T = 12$ K and 200 K) along the [111] direction are shown in Fig.\ \ref{phonAnom}. Anomalies similar to those first reported by Mook and Nicklow\cite{Mook79} are observed for \textit{all} compositions $x \neq 0$. As expected, the phonon softening is particularly strong in the crystal with the largest yttrium concentration ($x = 0.33$) and at the highest experimental temperature. However, it is remarkable that the effect is still present at temperatures and Y concentrations where the material is unambiguously in the $B$ phase. Upon decreasing temperature and/or Y concentration, i.e. as Sm gets closer to divalent, the shape of the dispersion changes and the softening becomes less pronounced. In the \SmYS{0.83}{0.17} compound, where the Sm valence is almost temperature independent, the LA phonon dispersion does not change with temperature within experimental accuracy. This further indicates that the phonon anomaly is directly connected with the valence instability.


\subsection{Magnetic excitation spectra in powder samples}
\label{subsec:tofResults}

 \begin{figure}
	\includegraphics [width=0.70\columnwidth, angle=-90] {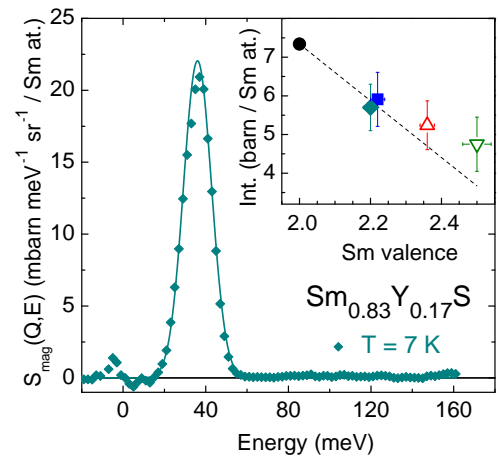}
\caption{\label{tof_7K} (Color online) Low-temperature ($T = 7$ K) magnetic scattering function in \SmYS{0.83}{0.17}, measured with an incident neutron energy of 300 meV at a scattering angle $\langle 2\theta \rangle = 5^{\circ}$ (momentum transfer $Q = 1.26$ \AA$^{-1}$ for $E = 36$ meV). The elastic peak, fitted using the spectral lineshape obtained from a vanadium standard, has been subtracted out. Inset: integrated intensities of the inelastic peaks at $T = 7$ K, normalized to the Sm concentration $x = 0$ (circle), 0.17 (diamond), 0.25 (square), 0.33 (triangle), and 0.45 (inverted triangle), plotted as a function of the Sm valence; dashed line represents the dependence expected for a weighted superposition of magnetic scattering from \Sm{2} and \Sm{3}.}
 \end{figure}

In order to extract the magnetic contribution from the experimental TOF data, one first needs to subtract out the phonon background, which can be estimated from data obtained in a separate measurement of La$_{0.75}$Y$_{0.25}$S (sample $e$). The correction procedure was described in Ref.\ \onlinecite{Murani94}, and validated by a number of subsequently studies.
The low-temperature ($T = 7$ K) magnetic response is quite similar for all Y concentrations. Results for $x = 0.17$ are displayed in Fig.\ \ref{tof_7K} over a wide range of energy transfers $-20 \leqslant E \leqslant 160$ meV. The spectrum essentially consists of a single, relatively broad, peak at about 34 meV, which can be ascribed to the \FF\ SO transition in Sm$^{2+}$, in agreement with earlier data.\cite{Mook78,Holl-Mor88} The integrated intensity of the peak, normalized to the concentration of Sm ions, decreases when $x$ increases (inset in Fig.\ \ref{tof_7K}) as a result of the enhanced trivalent character. However, the variation shown in the inset appears to deviate from a linear slope (dashed line) and to level off at the highest Y concentrations. 
The width of the \hbox{34-meV} peak is comprised between 16 and 18 meV for the different samples, which exceeds the experimental resolution ($\text{FWHM} \approx 11$ meV) by a factor of about 1.5. This broadening cannot be due to simple relaxation processes, since the peak shape is not Lorentzian. It can result from an internal structure of the peak, as observed in Ref.\ \onlinecite{Holl-Mor88}, and/or an energy dispersion smeared out by \hbox{$Q$-space} averaging in powder data. This interpretation was confirmed by our previous single-crystal study of \SmYS{0.83}{0.17}, \cite{Alekseev02_PRB} and is further substantiated by the new results to be presented in Section~\ref{subsec:taxResults}

Surprizingly, in spite of the strong mixed-valence character of the alloy systems, magnetic intensity could be detected, at low temperature, neither in the energy range of the Sm$^{3+}$ \HH\ inter-multiplet transition ($E^{3+}_{\text{so}} \sim130$ meV) nor in the quasielastic region. This result is important to understand the nature of the mixed-valence ground state in this system, and will be discussed in more detail hereafter.

 \begin{figure}
	\includegraphics [width=0.75\columnwidth, angle=-90] {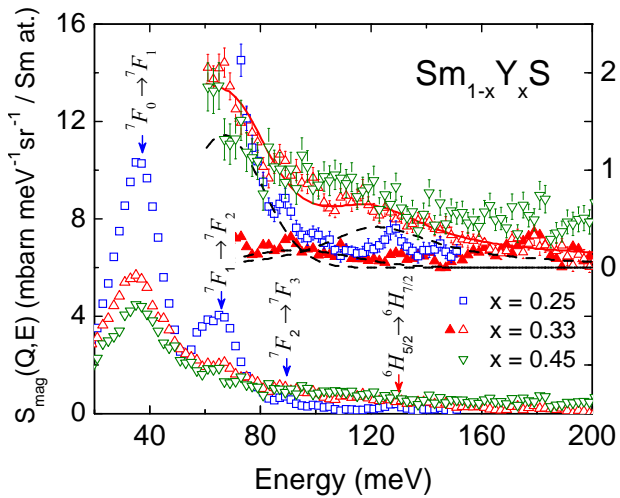}
\caption{\label{tof_fcn(T)} (Color online) Room-temperature (RT) magnetic scattering function in \SmYS{1-x}{x}, measured at a scattering angle $\langle 2\theta \rangle = 5^{\circ}$ (momentum transfer $Q = 1.26$ and 3.1 \AA$^{-1}$ for incident neutron energies $E_i = 36$ and 130 meV, respectively). Upper right: expanded plot on the same energy scale; data for $x = 0.33$ at $T = 7$ K have been added (closed triangles); solid line: fit of the RT spectrum for \SmYS{0.67}{0.33} using three peaks (dashed lines) associated with \Sm{2} ${^7F_i} \rightarrow {^7F_{i+1}}$ ($i = 1$, 2) (Gaussian), and \Sm{3} ${^6H_{5/2}} \rightarrow {^6H_{7/2}}$ (Lorentzian)  intermultiplet transitions.}
 \end{figure}

On increasing temperature, strong changes are observed in the magnetic spectral response. A representative set of spectra is collected in Fig.\ \ref{tof_fcn(T)}. Extra inelastic peaks at energies up to 90 meV, which occur in a similar way for all samples studied, can readily be ascribed to transitions from the Sm$^{2+}$ excited multiplets ($J = 1, 2$). Furthermore, in the spectra for $x = 0.33$ and 0.45, a very broad signal ($\text{FWHM} \sim 50$ meV), centered around $E \approx 130$ meV, appears at the highest experimental temperature ($T \approx 300$ K). Its energy roughly corresponds to the first SO transition in Sm$^{3+}$. Under the same conditions, a similar signal is not observed in samples with lower Y concentrations. This can be explained by the fact that the samples with $x  = 0.33$ and 0.45 are the most trivalent in the series studied, with properties indicative of the $G$ phase for temperatures higher than 200 K and 100 K, respectively. It seems therefore natural to observe a fingerprint of the Sm$^{3+}$ configuration in this regime. However we found no detectable intensity around 130 meV in the spectrum for \SmYS{0.55}{0.45} measured at $T = 200$ K ($G$ phase), even though the valence is higher than in \SmYS{0.67}{0.33} at room temperature. The reason for this discrepancy is not yet understood. Another intriguing issue is the absence of qualitative difference in the Sm$^{2+}$ component between the $B$ and $G$ phases. 

 \begin{figure}
	\includegraphics [width=0.75\columnwidth] {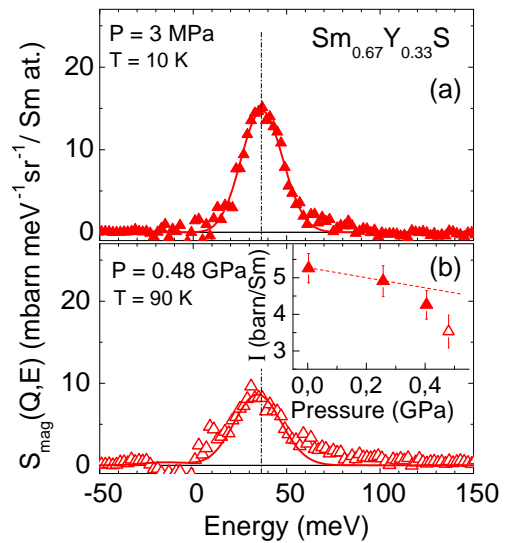}
\caption{\label{tof_hp} (Color online) Magnetic scattering function in \SmYS{0.67}{0.33}: (a) at $P = 3$ MPa ($T = 10$ K); (b) at $P = 0.48$ GPa ($T = 90$ K); solid lines are Gaussian fits to the spectra; dash-dotted line through the low-pressure maximum emphasizes the shift in the peak position. Inset in (b): pressure dependence of the integrated inelastic magnetic intensity.}
 \end{figure}

One possibility to stabilize the $G$ phase at low temperature within the present composition range, $0 \leqslant x \leqslant 0.45$, is to apply an external pressure of a few hundreds of megapascal. Sample $d$ ($x = 0.33$), which is close to the critical concentration, was selected for this experiment. The maximum pressure at $T_{\text{min}}$ that could be obtained in the He pressure cell was 0.4 GPa, which turned out to be insufficient to drive the sample into the $G$ phase. As can be seen from the comparison of integrated intensities for the \hbox{34-meV} peak measured under different conditions (inset in Fig.\ \ref{tof_hp}), the \hbox{$B$-$G$} transition could be reached by heating up to $T = 90$ K, because the melting of the He transmitting medium produced an increase of pressure to $P = 0.48$ GPa. The result was a sizable reduction in the peak intensity. Since the normal thermal depopulation of the Sm$^{2+}$ ground state at 90 K is estimated to be no more than 3 per cent, this reduction must reflect a significant decrease in the Sm valence. Another interesting observation is that the peak  shifts to lower energies by more than 3 meV. A similar, though much weaker, shift was also observed at ambient pressure with increasing Y concentration. The main effects of pressure are thus similar to those achieved by Y doping. However, the \HH\ SO transition could not be observed under pressure, the reason being probably its weak intensity and large linewidth, as well as the rather high background from the pressure cell.


\subsection{Single-crystal study}
\label{subsec:taxResults}
 
 \begin{figure}
	\includegraphics [width=0.70\columnwidth ] {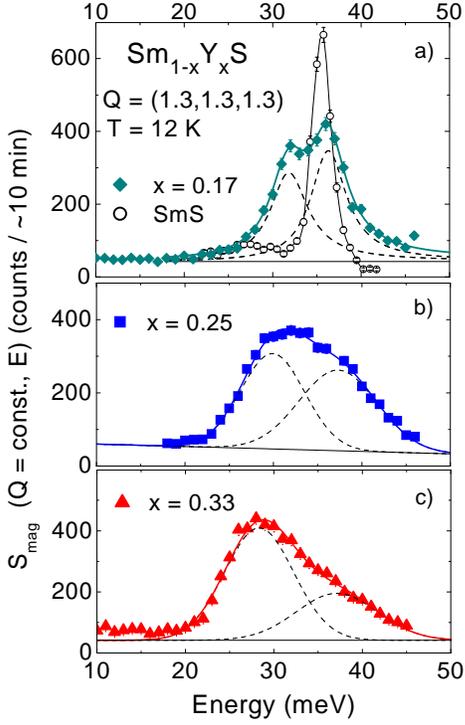}
\caption{\label{tax_spectra} (Color online)  Inelastic neutron spectra in \SmYS{1-x}{x} single crystals for $\bm{Q} = (1.3, 1.3, 1.3)$ at $T = 12$ K; lines are fits by two Gaussian peaks. The \FF excitation in pure SmS is shown as open circles in (a) (line is a guide to the eye).}
 \end{figure}

 \begin{figure}
	\includegraphics [width=0.75\columnwidth, angle=-90] {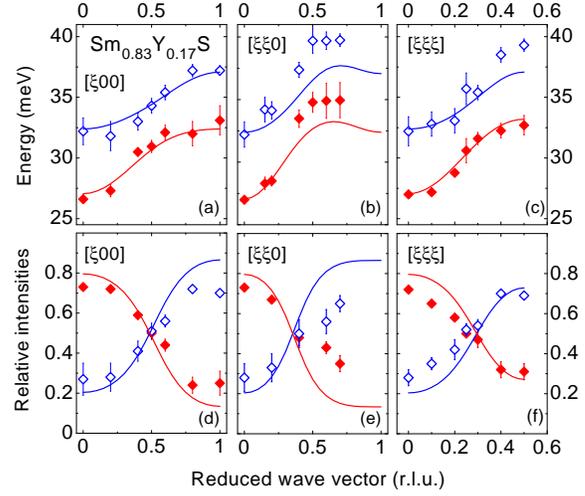}
\caption{\label{tax_disp13} (Color online) Wave-vector dependence of the energy (upper frames), and relative intensities (lower frames), of the two magnetic excitations in \SmYS{0.83}{0.17} measured along main symmetry directions at $T = 12$ K; lines represent the model MF-RPA calculation (see text)}
 \end{figure}

 \begin{figure}
	\includegraphics [width=0.70\columnwidth, angle=-90] {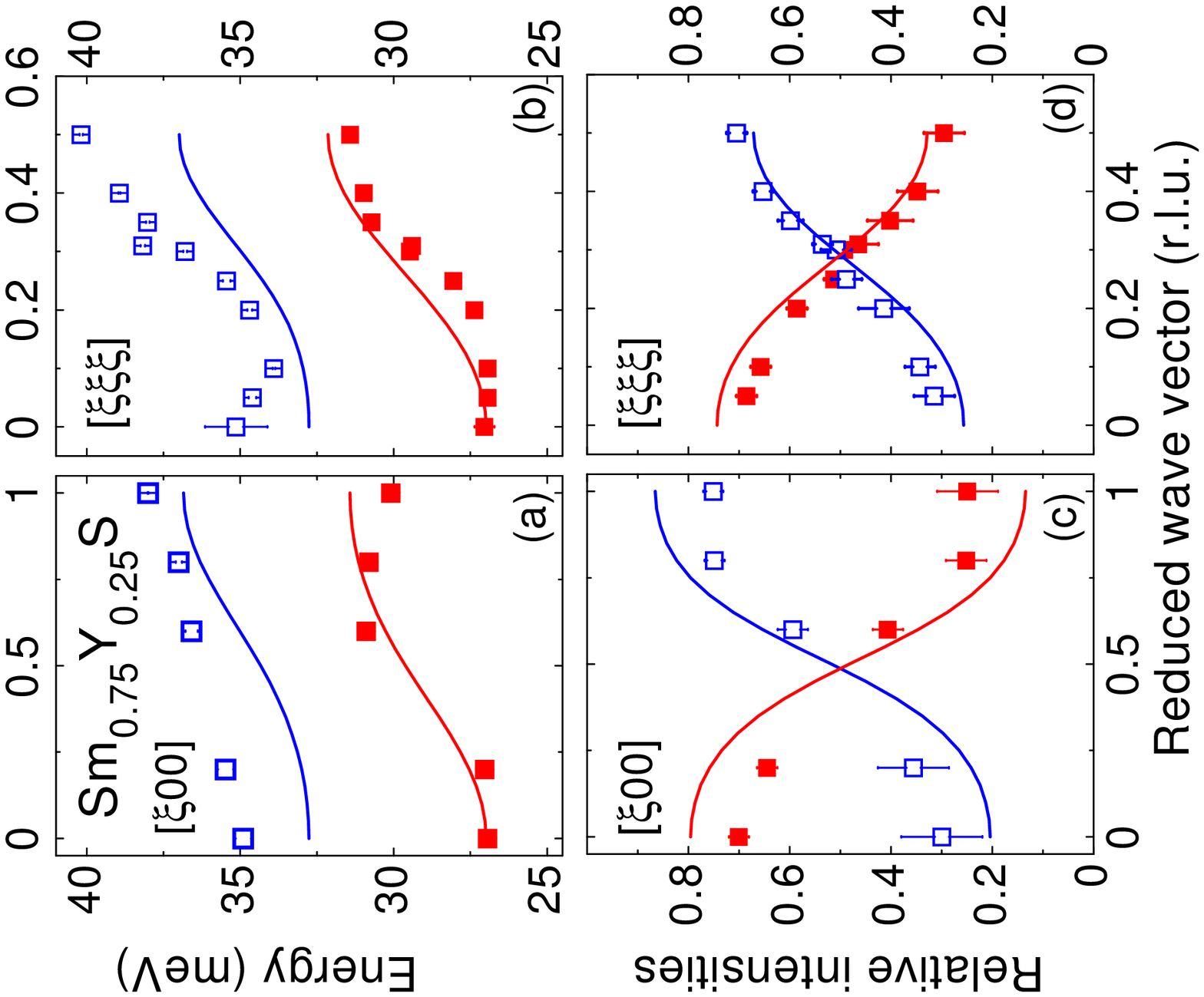}
\caption{\label{tax_disp25} (Color online) Same as Fig.\ \ref{tax_disp13} for \SmYS{0.75}{0.25}.}
 \end{figure}

 \begin{figure}
	\includegraphics [width=0.75\columnwidth, angle=-90] {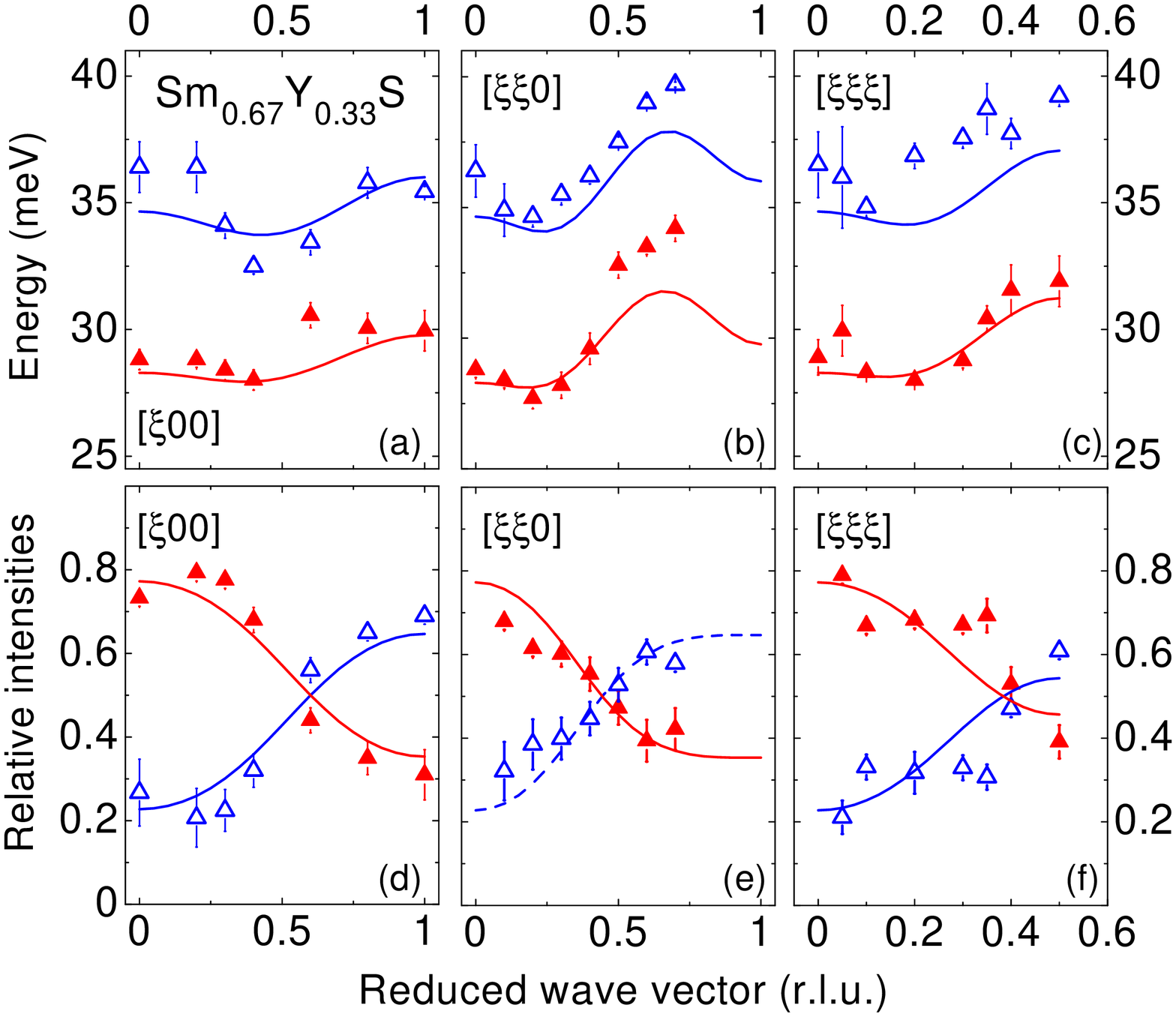}
\caption{\label{tax_disp33} (Color online) Same as Fig.\ \ref{tax_disp13} for \SmYS{0.67}{0.33}.}
 \end{figure}

Three different single-crystals with compositions $x = 0.17$ ($a$), 0.25 ($b$), and 0.33 ($c$) were studied in the triple-axis experiments. The general evolution of the magnetic
\footnote{In the TOF measurements, it was observed that the phonon contribution in the whole \SmYS{1-x}{x} series is very weak at low momentum transfer. Therefore, in the present triple-axis measurements with momentum transfers restricted to the vicinity of the first Brillouin zone, the phonon contribution was neglected and the measured signal treated as purely magnetic.}
spectral response in the range of the Sm$^{2+}$ \FF\ transition is summarized in Fig.\ \ref{tax_spectra}.  The most striking feature, already reported in Ref.\ \onlinecite{Alekseev02_PRB} for $x = 0.17$, is the double-peak structure, which is now confirmed to exist in the entire MV  \hbox{$B$-phase} regime. One can see in Fig.\ \ref{tax_spectra} that, for equivalent measuring conditions, increasing the Y concentration results in a transfer of spectral weight from the high-energy to the low-energy mode. The dispersion curves of the two magnetic branches along the main cubic symmetry directions are plotted in Figs.\ \ref{tax_disp13}, \ref{tax_disp25}, and \ref{tax_disp33}, together with the $Q$ dependences of the mode intensities. For each concentration, one notes the similarity of the energy dispersions for the two branches, and the exchange of intensity between them as a function of the reduced wave vector $q$, taking place along all three main symmetry directions. Increasing Y concentration results in $i$) a reduction of the energy dispersions for both modes, and a comparable change in their general behavior; $ii$) the transfer of part of the total intensity from the higher to the lower excitation; $iii$) an increase in the average splitting between the two branches (compare upper frames in Figs.~\ref{tax_disp13} and \ref{tax_disp33}). The analogy between the dispersions and $x$ dependences of the two modes points to a common origin; the fact that they exchange intensity further suggests that they may have the same symmetry.

 \begin{figure}
	\includegraphics [width=0.75\columnwidth] {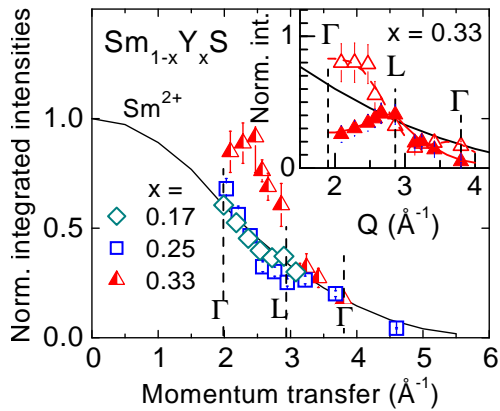}
\caption{\label{tax_Qdepdce} (Color online) $Q$ dependence of the total integrated intensity of the two magnetic excitations in \SmYS{1-x}{x} alloys at $T = 17$ K along the [111] direction; solid line: calculated form-factor for the \FF transition; $\Gamma$ denotes the Brillouin zone center, $L$ the zone boundary point $\bm{Q} = (\frac{3}{2},\frac{3}{2},\frac{3}{2})$. Intensities have been normalized at the highest experimental $Q$ value to the calculated \Sm{2} form factor. Inset: partial integrated intensities of the two magnetic excitations for $x = 0.33$, normalized at $Q = 2.66$ \AA$^{-1}$; open triangles: lower-energy peak; closed triangles: higher-energy peak; solid line: calculated form factor for the \FF\ transition; wiggling lines are guides to the eye denoting the oscillating behavior of partial intensities (see text).}
 \end{figure}

The integrated intensity associated with the magnetic double-peak structure is plotted in Fig.\ \ref{tax_Qdepdce} as a function of the momentum transfer $Q$  for different yttrium concentrations, $x = 0.17$, 0.25, and 0.33. The total intensity of the two modes decreases with increasing momentum transfer, confirming that their origin is magnetic. The existence of a superimposed modulation, also observed in pure SmS, with a period approximately matching the dimension of the Brillouin zone, indicates that they are not single-site. The overall $Q$ dependence appears to be steeper than the calculated single-ion form factor associated with the \FF\ excitation in Sm$^{2+}$, depicted by the solid line in the figure. These effects are most pronounced for $x = 0.33$. In the inset, the $Q$ dependences of the partial intensities for each magnetic mode have been plotted separately for the latter composition. They exhibit a quite remarkable oscillating behavior, again with a period equal to the size of Brillouin zone in the direction considered. The intensity of the low-energy component, averaged over one period, falls off appreciably faster with increasing $Q$ than the single-ion form factor, and makes the dominant contribution to the observed decrease of the total intensity. This gives us a hint that the electron states involved in this excitation may have a spatial extension larger that the original $4f$ shell.

\begin{figure}
	\includegraphics [width=0.70\columnwidth] {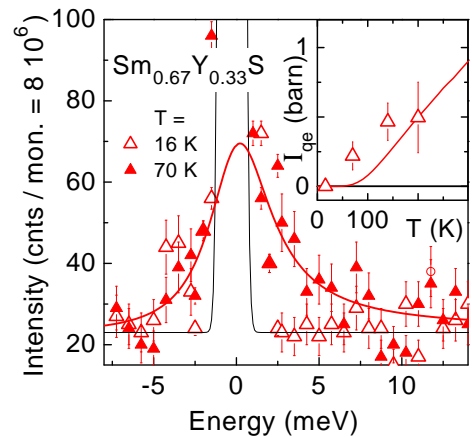}
\caption{\label{tax_qelast} (Color online) Quasielastic neutron scattering spectra for $\bm{Q} = (1.45, 1.45, 1.45)$ in single-crystal \SmYS{0.67}{0.33} at $T = 16$ and 70 K. The spectrum for $T = 70$ K was fitted by a quasielastic Lorentzian lineshape with $\Gamma \approx 6$ meV; solid line: fit of the elastic peak at $T = 70$ K. Inset: temperature dependence of the integrated intensity of the quasielastic signal; line: calculated temperature dependence from thermal population effects on the $^7F_0$, $^7F_1$, and $^7F_2$ multiplets for ionic \Sm{2}, taking into account the \hbox{$T$-induced} valence change; experimental values have been normalized to the calculated intensities at $T = 200$ K.}
 \end{figure}

Another important point concerns the temperature evolution of the quasielastic magnetic signal. The most detailed data set has been obtained for the composition $x = 0.33$ (Fig.\ \ref{tax_qelast}). In this compound, quasielastic scattering cannot be detected in the spectrum measured at $T = 16$ K, but a Lorentzian peak, centered at $E = 0$, is already recovered at  $T = 70$ K. Its spectral weight then grows steadily up to room temperature, as shown in the inset, while its width increases from $\sim6$ meV at 70 K to $\sim12$ meV at 200 K. It is quite notable in the plot that the appearance of the quasielastic signal takes place at a temperature much lower than expected from the normal thermal population of \Sm{2} excited SO multiplets (shown as a solid line), suggesting that some renormalization mechanism, associated with the valence fluctuation, modifies the low-energy electron states within this temperature range. 

\begin{figure}
	\includegraphics [width=0.60\columnwidth, angle=-90] {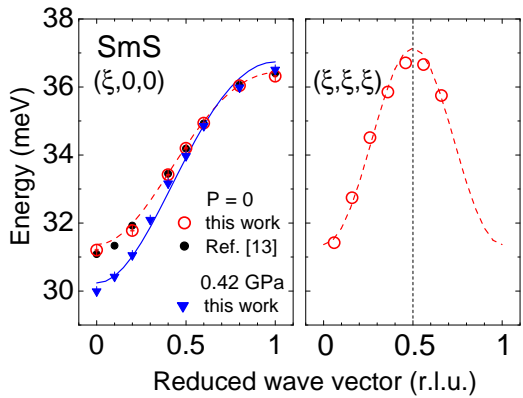}
\caption{\label{tax_hp} (Color online) Energy dispersion of the magnetic SO excitation in SmS along two main symmetry directions [100] and [111]; open circles: $P = 0$, $T=15$ K; triangles: $P = 0.42$ GPa, $T= 80$ K; closed circles: ambient pressure data at $T = 77$ K from Ref.\ \onlinecite{Shapiro75}; dashed lines represent the calculated energy dispersions using exchange parameters from Table \ref{tabFitParam}.}
 \end{figure}

Finally, we have studied the effect of a moderate hydrostatic pressure on the magnetic excitation spectrum in pure SmS. The dispersion curves of the SO excitation measured at ambient pressure and for $T = 15$ K are plotted in Fig.\ \ref{tax_hp} for the directions [100] and [111]. The [100] data are in excellent agreement with the earlier results of Shapiro \etal.\cite{Shapiro75} The dispersion along [111], which was not studied in the latter work, can be very well reproduced using the values of exchange parameters obtained in Ref.~\onlinecite{Shapiro75} from a refinement of the [100] and [110] data. The energy width of the experimental peak is found to be resolution limited.  At the maximum pressure of 0.42 GPa accessible with the He-gas pressure cell, SmS is still in the $B$ phase. As could be expected, no dramatic change in the magnetic excitation spectrum occurs with respect to $P = 0$ since, in the pure compound, the divalent state is retained up to the first-order \hbox{$B$-$G$} transition. However,
measurements performed along the [100] direction at $T = 80$ K (where only the SO ground state is significantly populated) indicate a sizable enhancement of the dispersion (Fig.\ \ref{tax_hp}, left frame).

\section{Discussion}
\label{sec:discussion}


\subsection{Lattice properties and valence}
\label{subsec:lattProp}

The XAS data reported in Fig.~\ref{valenceThermexp}(a) for samples $c$ and $d$ and in Ref.\ \onlinecite{Alekseev02_Pha} for the lower Y concentrations clearly indicate that the Sm valence is nonintegral not only in the $G$ phase (samples $c$ at $T > 200$ K and $d$ at $T > 100$ K), but also in the $B$ phase (samples $c$ and $d$ at low temperature, and $a$ and $b$ throughout). This result is in line with the conclusion reached by Tao and Holtzberg in their early work on \SmYS{1-x}{x} solid solutions.\cite{Tao75} Whereas the nature of the valence mixing occurring in the $B$ phase remains an open question, the softening of the LA phonon dispersion observed in  the present study (Fig.\ \ref{phonAnom}) lends support to the contention that this mixing is homogeneous in nature.

Another interesting issue is the strong lattice expansion that takes place on cooling in the $x = 0.33$ and 0.45 alloys (Fig.~\ref{valenceThermexp}(a)), giving rise to a pronounced negative peak in the $T$ dependence of the linear thermal expansion coefficient $\alpha$ (Fig.~\ref{valenceThermexp}(b)). These effects indicate a strong response of the lattice to the valence transition, with the position of the minimum approximately corresponding to the transition temperature for each composition. On the other hand, despite the significant deviation of their valence from 2+, the $B$-phase alloys exhibit a normal, positive, thermal expansion coefficient, with a $T$ dependence similar to that in La$_{0.75}$Y$_{0.25}$S. Qualitatively, the observed behavior can be explained by the larger ionic radius of \Sm{2}, whose relative fraction increases in the average at low temperature. Using the values of the valence plotted in Fig.~\ref{valenceThermexp}(a), a simple linear interpolation (``Vegard's law'') predicts a volume increase by 1.4 \% and 1.5 \% for $x = 0.33$ and 0.45, respectively, between 300 K and 10 K, to be compared with experimental values of 2.3 \% and 1.3 \%. Aside from possible uncertainties in the above estimates, the difference might point to some additional mechanism contributing to the thermal expansion. In the case of pure SmS, high-pressure experiments \cite{Matsuba04,Iwasa05} have shown the existence, for pressures comprised between $\sim 0.4$ and 1.2 GPa, of a negative peak in $\alpha (T)$ whose magnitude is comparable to that found in the present work. It can be noted, however, that, unlike the lattice constant curves displayed in Fig.~\ref{valenceThermexp}(a), those reported for SmS under pressure show a normal decrease on cooling below room temperature followed by a minimum around $\sim 100$ K. As a result, the anomalous dependence, with negative values of $\alpha$, develops only at lower temperatures.\cite{Matsuba04} In Refs.~\onlinecite{Matsuba04} and \onlinecite{Iwasa05}, the results were fitted to Schottky anomalies, assuming a two-level state with an energy gap decreasing gradually from 80--100 K to $\sim 30$ K over the experimental pressure range. It is presently not clear to what extent such a picture would be applicable to the \SmYS{x}{1-x} alloys because, unlike pure SmS at high pressure, these systems exhibit a clear valence transition as a function of temperature whose effect on the lattice constant certainly cannot be dismissed.

The present results can be compared with those obtained previously on SmB$_{6}$ and its alloys with La, which also exhibit large negative peaks in their thermal expansion coefficients. By simultaneously analyzing the temperature dependences of the heat capacity, $C_P$, and the thermal expansion, one can identify three main terms (lattice excitations, $f$-electron magnetic excitations, gap in the electronic density of states), each one being assigned a particular Gr\"uneisen coefficient\cite{Nefe03} and contributing independently (in first approximation) to both $C_P$ and $\alpha$. In pure  SmB$_{6}$, it turns out that the first two terms dominate the thermal expansion respectively above and below 150 K, whereas the third one is significant only below 50 K. The important point is that the amplitude of the minimum never exceeds $-7  \times 10^{-6}$~K$^{-1}$, as compared to $-5.5  \times 10^{-5}$~K$^{-1}$ in the present compounds. This difference by almost an order of magnitude reinforces the assumption that the conversion of the Sm valence state from 3+ to 2+ is the leading mechanism in \SmYS{x}{1-x} alloys undergoing a valence transition as a function of temperature, with other thermodynamic effects associated with either electronic or lattice degrees of freedom making only minor contributions. This fact should be kept in mind when comparing sulfide and hexaboride MV Sm compounds.

As to the Sm mean square displacement $\langle u^2_{\mathrm{Sm}} \rangle ^{\ast}(T)$, its non-monotonic variation in \SmYS{0.67}{0.33}, with a minimum occurring at about the valence transition temperature (Fig.~\ref{valenceThermexp}(c)), seems to be the consequence of a change in the inter-atomic interaction potential due to the \BG\ transition. Such a minimum was not observed, down to 100 K, in an earlier study by Dernier et al.\cite{Dernier76} on an alloy with $x = 0.3$, and $\langle u^2_{\mathrm{Sm}} \rangle ^{\ast}(T)$ was then found to increase monotonically with temperature between 100 and 300 K. However, in view of the lattice constant data for the same alloy reported in Ref.~\onlinecite{Dernier76}, its composition seems to correspond, in our case, to a higher Y content, close to that of sample $d$. If so, the minimum would have been expected to occur near 100 K or below, and thus might have been missed in the previous measurements. Let us mention for completeness that the sulphur mean square displacement $\langle u^2_{\mathrm{S}} \rangle ^{\ast}(T)$ could not be determined in the present experiments, but was found in Ref.~\onlinecite{Dernier76} to go through a minimum around 230 K. The interpretation proposed by Dernier et al. was based on a breathing-shell model in which Sm valence fluctuations couple to phonon modes having a component of an isotropic deformation of the nearest-neighbor S octahedron. It was predicted that the only effect giving rise to significant spectral weight would consist in a softening of LO modes. The present results suggest that the situation is indeed more complicated and that acoustic modes also may be significantly affected. This may provide a basis for further theoretical work on the lattice dynamics of MV Sm compounds.


\subsection{Magnetic dynamics}
\label{subsec:magnDyn}

\subsubsection{Quasielastic scattering}

From the neutron scattering results presented in the preceding section, the magnetic spectral response of the \SmYS{1-x}{x} alloys was shown to consists of three main components: $i$) a quasielastic signal which exists only for temperatures higher that $\sim 50$ K; $ii$) a doublet of peaks in the energy range expected for the Sm$^{2+}$ \FF\ inter-multiplet transition (with extra peaks attributable to transitions from excited multiplets appearing at increasing temperature); $iii$) only in the $G$ phase and at high enough temperature: a very broad signal around 130 meV, which can be traced back to the Sm$^{3+}$ \HH\ transition.

The temperature dependence of the quasielastic signal for $x = 0.33$, summarized in Fig.\ \ref{tax_qelast}, corresponds to a temperature range below the transition, which occurs at $\approx 160$--180 K. The system is thus expected to be predominantly in the $B$ phase. The variation of the intensity shown in the inset turns out to be in good agreement with the TOF results (50 K $< T < 250$ K) previously reported by Weber \etal\cite{Weber89} for a sample with $x=0.25$. In the latter study, the \BG\ transition presumably took place around 150 K but no clear effect was noted on the neutron data.
By extending the range of measurements to $T = 16$ K, we were able to confirm the complete suppression of the quasielastic component at low temperature. Furthermore, the reappearance of a sizable signal already at 70 K cannot be explained by a simple ``superposition'' picture, in which Sm$^{2+}$ and Sm$^{3+}$ configurations would contribute independently. Indeed, there is no scattering channel available for quasielastic scattering in the $J = 0$ ground state of Sm$^{2+}$, and the thermal population of the excited $J = 1$ multiplet at $T = 70$ K is too small to explain the intensity observed experimentally. In the data of Ref.\ \onlinecite{Weber89}, measured in absolute units, it could already be seen that the signal measured exceeded the scattering cross section expected from Sm$^{2+}$ alone. The $J = 5/2$ multiplet of Sm$^{3+}$ could in principle provide an additional contribution, but the relatively small linewidth observed in Fig.\ \ref{tax_qelast} seems incompatible with the damping of the \HH\ transition, which is strong enough to completely wipe out the corresponding \hbox{130-meV} peak at $T = 70$ K. Furthermore, if this were the case, one would expect the signal existing at 70 K to remain visible down to 16 K, which is obviously not true experimentally. This leeds us to envision a spin-gaplike picture, similar to that proposed previously for SmB$_6$,\cite{Alekseev95_JETP} in which the ground state is a many-body singlet, different from the single-ion $^7F_0$ singlet of Sm$^{2+}$, and forming at temperatures below $\sim 50$ K. Quasielastic scattering then gradually reappears with increasing temperature as a result of low-lying excitations from this ground state.

\subsubsection{Inter-multiplet transitions}

One remarkable feature of the magnetic response of the \SmYS{1-x}{x} solid solutions is that, whereas we have found the low-energy part of the magnetic response to be deeply altered as a result of valence mixing, the inter-multiplet transitions corresponding to Sm$^{2+}$ are observed at about the same energies  ($\approx 35 $ meV for \FF) as in divalent compounds, notably SmS itself, and with the same type of energy dispersion as in the latter compound. There is, however, one major difference, which is the appearance of a second excitation located about 5 meV lower in energy than the original peak. Already in the early TOF experiments of Holland-Moritz \etal\cite{Holl-Mor83} on \SmYS{0.75}{0.25} powder, indication was found for a structure in the inelastic neutron peak, implying the existence of two or more distinct modes. This effect was tentatively ascribed to a magnetoelastic effect, assuming a coupling to occur between the dispersive \Sm{2} ``spin-orbit exciton'' branch and an optic phonon branch of the same symmetry. Evidence against this interpretation now comes from the observation that this doublet peak structure does occur in the compound with $x = 0.17$ even though, for this composition, the frequencies of the optic phonons lie well below the spin-orbit exciton band, making the suggested electron-phonon process basically ineffective. Furthermore, the $Q$-dependence of the intensities for the two peaks (taken either separately or as a whole) globally follows that expected for a pure magnetic process (single-ion form factor), ruling out a significant admixture of phonon character.

Another possible mechanism which might account for the double peak structure is a crystal-field (CF) splitting of the $J = 1$ \Sm{2} multiplet.\cite{Fishman02} Formally, the local symmetry lowering expected to result from the substitution of divalent Sm ions by Y$^{3+}$ on the first rare-earth coordination sphere can produce non-cubic components in the CF potential, which ought to result in a splitting of the $^7F_1$ excited triplet state. For instance, a tetragonal configuration of the charge environment will lead to a splitting of this triplet into two substates: $J_z = 0$ and $J_z = \pm 1$. However, this mechanism cannot explain the two-peak profile observed in the magnetic excitation spectra. The point is that, for all Y concentrations studied, there is no dominant configuration of the first rare-earth coordination sphere around each \Sm{2}. Even for $x = 0.17$,  the probabilities to find configurations with one, two, and three Y ions among the 12 nearest neighbors are 26, 30, and 20 per cent, respectively. Each of these configurations, which in total represent about three quarters of the sites and should thus dominate the magnetic response, gives rise to \textit{different} local distortions of the CF potential depending on how the Y ions are distributed amongst the 12 available sites. From such a picture, the occurrence of two well defined peaks seems very unlikely. This argument is further supported by previous neutron measurements \cite{Alekseev00} on the inhomogeneous MV compound Sm$_3$Te$_4$, in which inter-atomic distances are close to those in SmS. In that compound, the ratio of \Sm{2} to \Sm{3} is exactly one half, and the probabilities of finding 2+ and 3+ ions on the first rare-earth coordination sphere of Sm  are comparable. Nevertheless, the peaks in the neutron spectra are well defined and corresponds to the CF scheme corresponding to the original, undistorted, local symmetry of the rare-earth site with chalcogen atoms as the first nearest neighbors. The only indication of an effect of the CF distortion due to charge inhomogeneity was possibly a weak broadening of the peaks in the neutron spectra. Therefore, it is reasonable to assume that, in \SmYS{1-x}{x} too, the effect of non-cubic components of the CF potential is relatively unimportant, and can at most produce some broadening in existing SO excitations.

Other peculiarities of the extra mode, in particular the steep $Q$ dependence of its intensity displayed in Fig.\ \ref{tax_Qdepdce} or the energy dispersion, indicative of a coherent effect and thus a priori difficult to reconcile with a random local environment, are also against a CF origin. But at the same time, they might provide some clues as to what could be a more plausible mechanism to account for the neutron data. In Ref.\ \onlinecite{Alekseev02_PRB},  a simple phenomenological model was proposed, based on the assumption of two magnetic modes hybridizing with each other, whose dispersions can be represented in the formalism of a singlet-triplet mean-field--random-phase-approximation (MF-RPA) model. In the context of the``local-bound-state'' picture of Ref.\ \onlinecite{Kikoin95}, these modes are assigned respectively to the ``parent'' \Sm{2} \FF\ spin-orbit transition, and to a renormalized excitation ($J^{\ast} = 0  \rightarrow J^{\ast} =1$) associated with the quantum-mechanical MV state. Other physical origins of the second mode are in principle compatible with the model, provided it retains the right symmetry to hybridize with the spin-orbit exciton, but the present assumption provides a natural explanation for the steeper $Q$ dependence of the intensity because the local bound state is less localized than ionic $4f$ states. Let us note, however, that in contrast to the case of SmB$_6$, the intensities of both branches do not follow the variation of the (local or extended) single-site form factor because of exchange coupling and mode-mode interaction.

Following Ref. \ \onlinecite{Shapiro75}, the dispersions of the two magnetic modes ($\lambda = \alpha, \beta$ for the ``parent'' and renormalized spin-orbit excitations, respectively) in the noninteracting limit are given by

\begin{equation}
\hbar \omega _\lambda  (\mathbf{q}) = \Delta _\lambda  \left[ 1 - 2M^2_\lambda R(T) \frac{J(\mathbf{q})} {\Delta _\lambda} \right]^{1/2},
\end{equation}

where $ M^2_{\lambda}$ is the squared matrix element of the transition, $J(\mathbf{q})$ the Fourier transform of the Sm-Sm indirect exchange interaction, and $R(T)$ a temperature factor. The two modes interact with each other owing to \hbox{$f-d$} mixing or partial delocalization of $f$ electron, which is also responsible for the existence of the lower mode. Therefore mode-mode interaction can be regarded as the consequence of two effects: MV state formation (on-site effect) and \hbox{$f-f$} exchange interaction (inter-site effect). The mixing interaction between the two modes is thus written as the sum of a (dominant) on-site term $V_0$ and a (smaller) hybridization term $V_1(\mathbf{q}) = \sum\limits_j {V_{NN} \exp \left[ {i \mathbf{q} \left(\mathbf{r}_j - \mathbf{r}_0\right)} \right]}$ due to the hybridization ($V_{NN}$) of a mode from a given Sm at site $\mathbf{r}_0$ with the other mode from a second Sm at site $\mathbf{r}_j$ belonging to its first coordination sphere. The above interaction can be treated in standard second-order perturbation theory, yielding the energies and intensities for both hybridized modes.


\begin{table*}
\caption{\label{tabFitParam}Parameters of the model used to describe the dispersive magnetic modes in \SmYS{1-x}{x}. The indices $\alpha$ and $\beta$ refer to the two different excitations observed experimentally (see text).}
\begin{ruledtabular}
\begin{tabular}{l c r l r c c c c c c c c c}
$x$&Sm&\multicolumn{3}{c}{Number of}&\multicolumn{2}{c}{Energy}&\multicolumn{3}{c}{Exchange constants}&\multicolumn{2}{c}{Mixing parameters}&\multicolumn{2}{c}{Relative intensities}\\

&valence&\multicolumn{3}{c}{1st neighbors}&\multicolumn{2}{c}{(meV)}&\multicolumn{3}{c}{(meV)}&\multicolumn{2}{c}{(meV)}&\multicolumn{2}{c}{of bare excitations}\\

&&$n_1$&$n_2$&$n_3$&$E_{\alpha}$&$E_{\beta}$&$J_1$&$J_2$&$J_3$&$V_0$&$V_1$&$A_{\alpha}$&$A_{\beta}$\\  

\colrule
0\footnote{Ref.\ \onlinecite{Shapiro75}}&$2.00$&$12$&$6$&$24$&$36.2$&-&$0.043$&$0.025$&$-0.003$&-&-&-&-\\

0&$2.00$&$12$&$6$&$24$&$36.2$&-&$0.043$&$0.020$&$-0.003$&-&-&-&-\\
0.17&$2.20$&$10$&$4.95$&$20$&$36.2$&$32.5$&$0.050$&$0.020$&$-0.003$&$-0.60$&$0.25$&$0.58$&$0.42$\\
0.25&$2.22$&$9$&$4.50$&$18$&$36.2$&$31.5$&$0.054$&$0.020$&$-0.006$&$-0.60$&$0.25$&$0.55$&$0.45$\\
0.33&$2.34$&$8$&$4$&$16$&$36.2$&$30.5$&$0.056$&$0.025$&$-0.020$&$-0.60$&$0.25$&$0.44$&$0.56$\\
\end{tabular}
\end{ruledtabular}
\end{table*}


In Ref.\ \onlinecite{Alekseev02_PRB} the experimental dispersion curves for \SmYS{0.83}{0.17} for the the two modes were fitted using only two adjustable parameters $V_0$ and $V_1$. Interestingly, it was found that using the parameters derived from the fit of the energy dispersions, the above phenomenological model also qualitatively explains the transfer of intensity between the two modes as observed experimentally. In fact, the exchange interaction alone is sufficient to roughly reproduce the shape of the dispersion curves but the hybridization term is necessary to account for the variation of the intensities. In the calculation of $J(\mathbf{q})$, the exchange constants up to third nearest neighbors were considered following Ref.\ \onlinecite{Shapiro75}, and their values were assumed to be the same for both modes, equal to those previously determined in undoped SmS. $\Delta_{\alpha}$ was taken equal  to 36.2 meV, the single-ion energy of the \Sm{2} \FF\ transition, and $\Delta_{\beta}$ to 32.5 meV, from the empirical relation found in the Sm$_{1-x}$La$_{x}$B$_6$ series of alloys between the local bound state energy and the Sm valence.\cite{Alekseev97} 
In order to consistently analyze the data from all three samples, the fitting procedure applied previously \cite{Alekseev02_PRB} was modified so as to to take into account the general trend of $q$ dependencies for both energies and intensities in the entire series. This leads to the fitting curves drawn in Figs.\ \ref{tax_disp13}--\ref{tax_disp33}, with parameter values listed in Table \ref{tabFitParam}.

If we first consider the variation of exchange interactions, it appears that, as the concentration of nonmagnetic substituent increases, with an attending decrease in the lattice spacing, the antiferromagnetic contribution rises substantially (almost by a factor of 7), producing a shift of the minimum in the energy dispersion curves from $q = 0$ to a finite $q$ value, as can be seen in the figures. It is worth noting that the application of an external pressure to pure semiconducting SmS results in just the opposite effect on the exchange parameters, namely an enhancement of the ferromagnetic character in the dispersion curves. In Fig\ \ref{tax_hp}, the energy at the $\Gamma$ point (Brillouin zone center) appears to be lowered by approximately 1 meV at a pressure of 0.42 GPa. This corresponds to an increase of the Sm-Sm exchange coupling by more than 10 per cent. The effect of the mere reduction in the lattice spacing, calculated using published SmS compressibility data, \cite{Melcher75} is found to be only about 2 per cent. 
Therefore the enhancement of the interaction between Sm ions (accounted for by the exchange integrals) could be supposed to take place under the applied pressure, in analogy with a Eu chalcogenides.\cite{Gonch98} This fact indicate some differences existing between effects of physical and chemical pressure on the one hand, and between inter-ion interaction in normal and mixed valence state on the other one.

Regarding hybridization parameters (normalized intensities of the bare excitations 
$A^2_\lambda = {{M^2_\lambda} / {\left( M^2_\alpha + M^2_\beta \right)}}$
and mode-mode interaction $V(q)$), the variation with Y doping cannot be uniquely ascribed to one particular parameter without making an extra assumption. In the present analysis, it was supposed that the main effect is a change in the relative weight ($A_{\beta}/ A_{\alpha}$) of the two modes, while other parameters ($V_0$, $V_1$) were fixed at the same values as in Ref.\ \onlinecite{Alekseev02_PRB}  for $x = 0.17$. The results reported in Table \ref{tabFitParam} indicate a significant enhancement of the lower ``renormalized'' excitation. It can also be noted that, in \SmYS{1-x}{x}, the renormalization of the energy of the lower peak as valence increases is not so strong as for the ``exciton" peak in Sm$_{1-x}$La$_x$B$_6$. Whereas for \SmYS{0.83}{0.17} an optimum fit is achieved using the value of $\Delta_{\beta}$ calculated from the empirical relationship derived in hexaborides \cite{Alekseev95_JETP} 
between the peak energy and the valence, the change in this value required to fit the data for larger concentrations $x = 0.25$ and 0.33, is not as large as would be expected from the latter relationship.

The above description of the single-crystal data is in a good agreement with the time-of flight data. Specifically, it gives us clues to some peculiarities of the magnetic spectral response measured on powder, like for instance the shift of the centroid of the 34 meV peak produced by Y doping or, even more clearly, by the application of pressure (see Fig\ \ref{tof_hp}). This effect can now be traced back to the enhancement of the low-energy subpeak, corresponding to the renormalized spin-orbit transition. The doping dependence of the integrated peak intensity (inset in Fig\ \ref{tof_7K}) can also be explained by the MV character of the extra magnetic mode. If the existence of this mode were simply due to a CF splitting of the original spin-orbit transition, this would result in a simple redistribution of the original \FF\ cross section between the two satellites, and the total integrated intensity should decrease proportionally to the deviation from the divalent state. As already noted, this is not verified experimentally. This discrepancy may be solved if one considers that the lower peak, which becomes stronger as the valence increases, has a different origin, with additional magnetic degrees of freedom getting involved as a result of the mixing with more extended electron states. 
Increase of the temperature is expected to suppress excitonic transitions in the spectrum quite fast. Indeed, Y doping at room temperature (or temperature increase for the particular Y concentration, x ( 0.33) does not result in some visible shift of this peak.


\section{Conclusion}

 \SmYS{1-x}{x} solid solutions with compositions $x = 0$, 0.17, 0.25, 0.33, and 0.45 have been studied by x-ray diffraction and inelastic neutron scattering, both on powder and single crystals. By relating lattice properties (thermal expansion coefficient $\alpha(T)$, mean square displacement $\langle u^2_{\text{Sm}} \rangle$) to the Sm valence obtained from XAS spectra, it was shown that all samples with $x \geq 0.17$ are in a homogeneous mixed-valence state. All measured samples are in the $B$ phase at low temperatures, and the $G$ phase is found only at $T \geq 200 K$ ($x = 0.33$) and $T \geq 100 K$ ($x = 0.45$). The \BG\ electronic phase transition is accompanied by a change in the inter-atomic interaction potential, thereby affecting the temperature dependence of $\alpha$ and $\langle u^2_{\text{Sm}} \rangle$. Further evidence for the anomalous character of the $B$ phase, even at the lowest Y content $x = 0.17$, is provided by the clear anomaly in the LA phonon dispersion curve (previously reported only for the $G$ phase).
 
For all concentrations studied, the magnetic excitation spectrum of powder samples displays a broad peak at energy $E \approx 34$ meV corresponding approximately to the \Sm{2} \FF\ spin-orbit transition. As Sm becomes more trivalent, this peak weakens and, at high enough temperatures and for valences exceeding $\sim 2.45$, a very broad signal, reminiscent of the \HH\ transition in \Sm{3}, is observed around 130 meV . This  strongly damped response likely indicates a very high rate of spin fluctuations. For $T \geq 70$ K, an additional quasielastic signal appears. As in SmB$_6$, its temperature dependence cannot be explained by simple thermal population effects, and can be traced back to the emergence of the quantum mechanical mixed-valence ground state below approximately this temperature.

The most remarkable feature of the magnetic spectral response of  \SmYS{1-x}{x} in the mixed valence regime was established by the single crystal experiments, namely an extra excitation occurring on the low-energy side of the \Sm{2} \FF\ spin-orbit transition. The evolution with increasing Sm valence of the intensities and energy dispersions of both spectral components can be successfully reproduced by a simple phenomenological model based on two hybridized modes. The formation of a loosely bound \hbox{$f$-electron} state (``MV exciton''), developed in detail for the case of mixed-valence SmB$_6$,\cite{Kikoin95} is proposed as the possible physical origin of the novel excitation.
The main difference with SmB$_6$ is the $q$ dependence of the energies and intensities for both mode. This effect is ascribed to \hbox{Sm-Sm} exchange interactions via the $d$ band, which is simultaneously involved in the process of \hbox{$f$-electron} delocalization due to valence instability. 

In conclusion, the present results shed some light on lattice and magnetic aspects of the long-debated problem of valence fluctuations in Sm systems. In particular, the INS results on magnetic dynamics reveals clear similarities, as well as conspicuous differences, between the sulfide alloys and the previously studied hexaboride compounds. In this connection, the discovery of properties reminiscent of the latter materials in a completely new family, Sm fulleride,\cite{Arvanitidis03} opens promising possibilities to check our present understanding of these phenomena.
\\

\begin{acknowledgments}
We thank M. Braden for scientific assistance and J. Dreyer and C. Goodway for technical support. We have benefited from fruitful discussions with K. A. Kikoin, A. S. Mishchenko, I. A. Smirnov, E. S. Clementyev, and R. S. Eccleston, The work was supported by the grant H\textcyr{SH}-2037.2003.2 and the RFBR grants No. 05-02-16426 and 05-02-16996. The hospitality and financial support of LLB (P. A. A.) and RAL (P. A. A., K. S. N. and N. N. T.) are gratefully acknowledged.
\end{acknowledgments}


\end{document}